\documentclass[onecolumn]{IEEEtran}
\usepackage{amsmath,amssymb,amsfonts,amsthm}
\usepackage{algorithmic}
\usepackage{algorithm}
\usepackage{array}
\usepackage[caption=false,font=normalsize,labelfont=sf,textfont=sf]{subfig}
\usepackage{textcomp}
\usepackage{stfloats}
\usepackage{url}
\usepackage{verbatim}
\usepackage{graphicx}
\usepackage[noadjust]{cite}
\usepackage{balance}
\usepackage{xcolor}
\graphicspath{ {./images/} }

\begin{document}
\title{\LARGE \bf
Blind System Identification in Linear Parameter-Varying Systems
}

\author{Javad Zahedi Moghaddam, Hamidreza Momeni, and Mojtaba Danesh
}

\maketitle

\begin{abstract}

Blind System Identification (BSI) is used to extract a system model whenever input data is not attainable.  Therefore, the input data and system model should be estimated simultaneously. Because of nonlinearities in a large number of systems, BSI problem is usually challenging to solve. In this paper, an innovative solution is proposed to deal with the BSI problem in nonlinear systems using the properties of the Linear Parameter-Varying (LPV) systems and Hidden Markov Models (HMM). More specifically, assuming the scheduling variable is not measurable, the dynamic of the LPV system is approximated. To solve the BSI problem in this context, the LPV structure is modeled as an HMM network and a modified Quasi-Static combination of Viterbi and Baum-Welch algorithms (QSVBW) is proposed to estimate the nonlinear mappings and scheduling variable signal. The applicability and performance of the suggested QSVBW algorithm have been justified by numerical studies.

\end{abstract}

\section{INTRODUCTION}\label{secI}

Modeling serves as the foundation for classification, clustering, and system identification tasks in statistical learning, enabling us to extract meaningful insights, make accurate predictions, and understand the underlying mechanisms governing the data or system of interest. Models in system identification help in understanding the dynamics, relationships, and parameters of the system under study \cite{c0_class_pedrycz2009experience,c0_class_karny2023model,moghaddam2022exact, c0_ident_risuleo2019modeling, c0_ident_danes2015enhanced,c1_ljung2008perspectives}. These models can be used for simulation, prediction, control, and optimization of the system. In some cases, not all the data required for system identification may be accessible. For example, certain inputs may not be observable. Therefore, standard system identification tools such as Prediction Error Methods (PEM) \cite{c1_ljung2008perspectives} cannot be applied, and specific methods including blind system identification (BSI) methods are needed to be employed \cite{c2_abed1997blind}. 

In BSI, both the system dynamics and some of the input signals are needed to be estimated. BSI is beneficial in many engineering areas where the required data for identification is incomplete. The BSI methods were used for the identification problem in image reconstruction \cite{c3_duan2016computed}, biomedical science \cite{c4_zhao2016blind}, and communications \cite{c5_mei2019blind} where the input data are not measurable. In linear and nonlinear system identification, some methods such as least square \cite{c6_identification1995least}, frequency domain decomposition \cite{c7_yao2018blind}, and Bayes approach \cite{c8_risuleo2019modeling} have been proposed to solve the BSI problems.
 
A large number of systems have nonlinear modes that linear models cannot catch, so in order to be able to approximate such models, experts prefer to use nonlinear models despite the challenges that come with nonlinear identification methods. Linear Parameter-Varying (LPV) models were introduced to endowed experts with a versatile configuration to estimate the nonlinear dynamics \cite{c9_toth2010modeling} in a linear structure. In the modeling process of LPV systems, the measurements of the three signals including input, scheduling variable, and output signal are required to identify the associated model of a system. In the LPV identification context, two main modeling approaches exist, input-output (LPV--IO) approach and state-space (LPV--SS) representation \cite{c9_toth2010modeling}. LPV--IO methods include data-driven techniques that are applied to identify an Input-Output model of a system in the LPV representation, while LPV-SS approach deals with the extraction of the state-space matrices in an LPV configuration that describe the dynamics of a system.

In this paper, the BSI problem is defined on an LPV system in order to utilize the properties of the LPV system, the Hidden Markov Models (HMM) are used as well to eventually design a versatile structure to solve the BSI problem. In other words, it is assumed that the input and output signals are measurable while the scheduling variable is not. To consider the most difficult case of identification with respect to the required data, LPV--IO based methods have been considered, and we call it blind LPV identification (BLPVI).

The rest of the paper is organized as follows: the LPV--IO model and discrete--time HMM are reviewed in section \ref{secII}. In section \ref{secIII}, the problem and the proposed approach are described comprehensively. Then, section \ref{secIV} has introduced the QSVBW algorithm to deal with the BLPVI problem. In section \ref{secV}, a numerical example is presented to examine the applicability and performance of the proposed method.  A discussion about the QSVBW algorithm is presented in section \ref{secVI}, and finally concluding remarks are given in section \ref{secVII}.

\section{Preliminaries}\label{secII}

\subsection{LPV-IO model}\label{secII.A}

LPV systems are an extension of LTI systems, where the dynamical relations between input and output signals are linear while the parameters of the model are assumed to be functions of a time-varying signal \cite{c9_toth2010modeling}. The modeling of LPV--IO systems has received a lot of attention in LPV identification literature.  An auto-regressive model with exogenous input (ARX) is the most common structure in LPV--IO modeling \cite{c9_toth2010modeling} which is described as
\begin{equation}
\label{eqn1}
y(k)=\sum_{i=1}^{n_{a}}a_{i}(p)y(k-i)+\sum_{j=0}^{n_{b}}b_{j}(p)u(k-j)+e(k),
\end{equation}
where $k \in \mathbb{Z}$  is the discrete time, $u:\mathbb{Z}\mapsto\mathbb{R}$, and $y:\mathbb{Z}\mapsto\mathbb{R}$ denote the input and output signals, respectively. $p::\mathbb{Z}\mapsto\mathbb{P}$ is the scheduling variable with the range $\mathbb{P}\subset \mathbb{R}^{n_p}$, and $e$ is an independent and identically distributed ({\it{iid}}) stationary white noise process. Scheduling variables designate the operational regimes of a system. The nonlinear coefficients $a(p)$ and $b(p)$ completely characterize an LPV--ARX model \cite{c9_toth2010modeling}. These nonlinear coefficients are divided into two categories based on their dependency on scheduling variables. The functional dependency of coefficients is assumed to be static if they depend on the current value of the scheduling variable. On the other hand, if the coefficients depend on the current and previous values of $p$, then the functional dependency is dynamic. In LPV--IO identification, the model of the system is directly identified from input-output ($IO$) data set. Using the linearity in the parameters of LPV--IO systems is a naive way to identification of the system \cite{c10_bachnas2014review}. This property allows us to use linear regression for the estimation of coefficient $a(p)$ and $b(p)$ under the assumption of additive white noise. Instrumental variable methods for the estimation of LPV--IO models under different noise conditions have been proposed \cite{c11_laurain2010refined}. In \cite{c11_laurain2010refined}, the output-error (OE) and Box-Jenkins types of noise models were used to identify the dynamics of the LPV--IO system. Recently, non-parametric methods such as Least Square Support Vector Machine (LS--SVM) \cite{c12_toth2011model}, and Bayesian Regression \cite{c13_golabi2017bayesian} have been introduced to reach an acceptable solution for this problem.

\subsection{Discrete--Time HMM}\label{secII.B}

In this section, the discrete-time Markov sequence and its theories are introduced, and then the discrete-time HMM is presented. A system is considered that its dynamics could be interpreted by $M$ distinct states at any time so that changes in states impress the behavior of the system. The corresponding state of the system changes based on a set of probabilities dependent on the state \cite{c14_nilsson2005first}. The time instance for a state change is denoted by $k$, and the corresponding state by $q(k)$. In a first-order Markov model, the transition probability of the state is not associated with the whole memory of the system, only the preceding state is taken into account \cite{c15_zucchini2009hidden}. This leads to the first-order Markov property \cite{c14_nilsson2005first} defined as
\setlength{\arraycolsep}{0.0em}
\begin{eqnarray}
\label{eqn2}
Pr\left( q(k)=s_{i}|q(k-1)=s_{j},\cdots, q(1)=s_{l} \right)&\nonumber\\
=Pr\left(q(k)=s_{i}|q(k-1)=s_{j}\right),&
\end{eqnarray}
\setlength{\arraycolsep}{5pt}
where ${\textbf S}=\{ s_1, \cdots, s_M\}$ is the discrete space of the states. The time-independency of the right hand of \eqref{eqn2} gives the state transition, $a_{ji}$, to be
\begin{equation}
\label{eqn3}
a_{ji}=Pr\left(q(k)=s_{i}|q(k-1)=s_{j}\right);1\leq i, j\leq M.
\end{equation}
According to standard probabilistic constraints \cite{c14_nilsson2005first}, the following properties should be applied to the state probabilities
\begin{equation}
\label{eqn4}
a_{ij}\geq0;\sum_{j=1}^{M}a_{ij}=1.
\end{equation}
In a Markov model, the transition probabilities for all states can be represented by a transition probability matrix
\begin{equation}
\label{eqn5}
{\mathbf A}=
 \begin {bmatrix}
a_{11} & \cdots &a_{1M}\\
\vdots & \ddots & \vdots\\
a_{M1} & \cdots & a_{MM}\\
\end{bmatrix},
\end{equation}
and the initial state probability for all states is expressed by
\begin{equation}
\label{eqn6}
\boldsymbol {\pi}=
\begin {bmatrix}
\pi_{1}&\pi_{2}&\cdots&\pi_{M}
\end{bmatrix}^{T},
\end{equation}
where
\begin{equation}
\label{eqn_def}
\pi_{j}=Pr\left(q(1)=s_{j}\right); 1\leq j\leq M.\nonumber
\end{equation}
Markov models are applied to the modeling of complex systems and networks such as economics models \cite{c16_rossi2006volatility}, and bioinformatics\cite{c17_collier2000extracting}.
The Markov model that was represented in the above paragraphs has some limitations and it could not be used in all situations. Hence, as an extension of the Markov model, HMM is introduced to reduce these restrictions. In an HMM, a distribution function is considered for each state, so each state produces an observation at time $k$, $O(k)$. For each state $j$ the distribution function is
\begin{equation}
\label{eqn7}
b_{j}\left(O(k)\right)=Pr\left(O(k)|q(k)=s_{j}\right).
\end{equation}

In this work, our observations are continuous, so we use the continuous probabilistic distributions. The continuous observation densities $b_{j}(O(k))$ in the HMM are created by using a parametric probability density function (pdf) or a mixture of several functions. The most utilized D-dimensional pdf is Gaussian, and it is as follows
\begin{equation}
\label{eqn8}
b_{jl}\left(O(k)\right)=\frac{exp\left(\left(O(k)-\mu_{jl}\right)^{T}\mathbf{\sigma}_{jl}^{-1}\left(O(k)-\mu_{jl}\right)\right)}{(2\pi)^{2} \mid{\mathbf{\sigma}_{jl}\mid}^{1/2}},
\end{equation}
where $\mu_{jl}$ and $\mathbf{\sigma}_{jl}$ are the mean and the covariance matrix of the $ l-th$ episode of the mixture density that is derived by corresponding state $j$.

A HMM like a first-order Markov model has a transition probability matrix $\textbf {A}$ and an initial state vector$\boldsymbol {\pi}$ that describe the HMM completely \cite{c15_zucchini2009hidden}. To apply an HMM on the different physical applications, one would face three challenging problems according to the configuration of the HMM \cite{c14_nilsson2005first}.
\begin{itemize}
\item Calculation the probability of the observation sequence, $ {\textbf O}=\begin {pmatrix}O(1)& O(2)& \cdots & O(k)\end{pmatrix}$, for a HMM that its parameters set, $\boldsymbol{\lambda}$, is given.  In other words, calculation of the likelihood function, $Pr(\mathbf{O}|\boldsymbol{\lambda})$, is desired. 
\item The problem is how to select the best state sequence, $ {\textbf q}=\begin {pmatrix} q(1)& q(2)& \cdots & q(k)\end{pmatrix}$, by considering the observation sequence, $ {\textbf O}$, and the HMM parameters $\boldsymbol{\lambda}$. In other words, the best-associated state sequence with the observation sequence and the model parameters is desired.
\item  Tuning the parameters of the HMM, $\boldsymbol{\lambda}$, so that the likelihood function, $Pr(\mathbf{O}|\boldsymbol{\lambda})$, is maximized.
\end{itemize}

The first problem is answered using the forward algorithm. In this algorithm, a forward variable is defined as follows
\begin{equation}
\label{eqn9}
\alpha_{i}(k)=Pr\left(\mathbf{O},q(k)=s_{i}|\boldsymbol{\lambda}\right),
\end{equation}
where $k$ is the time sample and $i$ is the corresponding state. To compute $Pr(\mathbf{O}|\boldsymbol{\lambda})$ in this algorithm, the key point is a recursive relation between the forward variables at sample $k$ and $k+1$. For more information about the forward algorithm see \cite{c14_nilsson2005first},\cite{c15_zucchini2009hidden}. In this work, we use such algorithms to compute the likelihood function. The second problem is answered using the Viterbi algorithm. Here, this algorithm is concisely explained for the next conclusions. Consider
\begin{equation}
\label{eqn10}
\delta_{i}(k)=\max_{q(1), q(2),\cdots,q(J-1)}\left[Pr(\mathbf{Q}_{J}(i)|\boldsymbol{\lambda})\right],
\end{equation}
where
\begin{equation}
\label{eqndef}
\mathbf{Q}_{J}(i)=\left(q(1),q(2),\cdots,q(J-1),q(J)=i,\mathbf{O}\right),\nonumber
\end{equation}
using mathematical induction, one can compute $\delta_{i}(k+1)$ as follows
\begin{equation}
\label{eqn11}
\delta_{i}(k+1)=b_{j}\left(O(k)\right)\max_{1\leq i \leq M}\left[\delta_{i}(k+1)\right].
\end{equation}
To retrieve the state sequence, it is necessary to keep track of the argument that maximizes \eqref{eqn11} for any state of $\mathbf{S}$, and this is done by saving the argument in an array.

The third problem is solved using the Baum-Welch algorithm. In this algorithm, an expectation maximization (EM) procedure is exploited to estimate the parameters of HMM \cite{c14_nilsson2005first}. In the Baum-Welch algorithm, the forward algorithm is used to compute some middle variables.

\section{Main}\label{secIII}

\begin{figure}
      \centering{}
      \includegraphics[trim=0cm 10cm 10cm 0cm,scale=0.35]{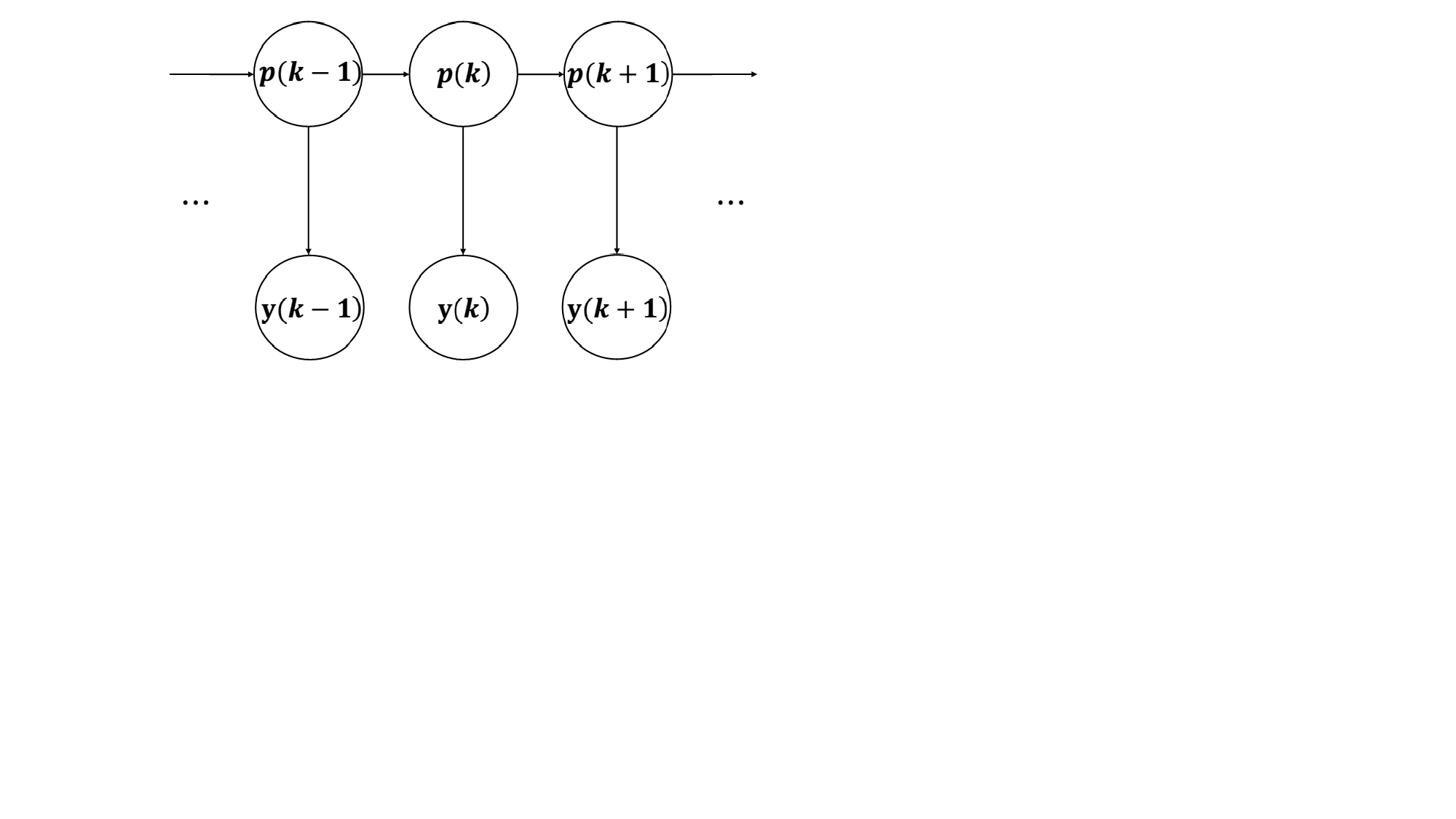}
      \caption{An HMM architecture of an LPV--IO model with static dependency on the scheduling variable}
      \label{figlabel1}
   \end{figure}

Here, for the sake of brevity, we consider an LPV-FIR model, i.e., an LPV-ARX model \eqref{eqn11} with $a_{i}=0, i=1,...,n_{a}$. The model of the system can be represented as
\begin{equation}
\label{eqn12}
y(k)=\sum_{i=1}^{n_{g}}g_{i}\left(p(k)\right)u(k-i+1)+e(k),
\end{equation}
where $n_b=n_g+1$, and we assume that the scheduling variable is static. The main idea for estimation of the scheduling variable is using an HMM with continuous observations so that the scheduling variable is considered as hidden states in an HMM that is shown in Fig. \ref{figlabel1}. In this work, we consider some following assumptions
\begin{itemize}

\item The functional dependency of coefficients is static.
\item The scheduling variable is a finite-amplitude step-wise signal, and the system is assumed to be single--input single--output (SISO).
\item The noise signal is white, and it is an {\it{iid}} and Gaussian process, that is, $e \sim \mathcal{N}(0,\sigma^{2})$.
\end{itemize}

Taking into account the above-mentioned assumptions, the output signal \eqref{eqn12} has a probability distribution at each sample
\begin{equation}
\label{eqn13}
y(k) \sim \mathcal{N}\left(\sum_{i=1}^{n_{g}}g_{i}(p(k))u(k-i+1),\sigma^{2}\right),
\end{equation}
where it can be considered as a Gaussian distribution requiring some of its parameters to be estimated. Since the input signal is known, the coefficients $g_i$s are the only parameters needed to be estimated. These coefficients depend on the scheduling variable; thus, the scheduling variable should be estimated first.

Based on the assumptions, the scheduling variable is chosen from a finite and discrete space, hence the amounts that it takes are treated as hidden states in an HMM.
We consider an HMM network similar to the network shown in Fig. \ref{figlabel1} that there are $N$ pairs of $I/O$ data and the value of the scheduling variable is chosen from $ \mathbf{\mathcal{P}}=\begin {Bmatrix} p_{1}& \cdots & p_{M} \end{Bmatrix} $ which is a discrete and finite space, then the complement likelihood function is as follows   
\setlength{\arraycolsep}{0.0em}
\begin{align}
    \label{eqn14}
Pr(\mathbf{Y},\mathbf{P}|\boldsymbol{\lambda})&=Pr(\mathbf{Y}|\mathbf{P},\boldsymbol{\lambda})Pr(\mathbf{P}|\boldsymbol{\lambda})\nonumber\\
&=\frac{\pi_{p(1)}}{{(2\pi {\sigma^2})}^{N/2}}  \prod_{k=1}^{N-1}a_{p_{k-1}^{k}}\prod_{k=1}^{N}exp\left(\frac{-\left({y(k)-\mu(k)}\right)^2}{2\sigma^{2}}\right)
\end{align}
\setlength{\arraycolsep}{5pt}
where
\begin{equation}
\label{eqn15}
\mu(k)= \sum_{i=1}^{n_{g}}g_{i}\left(p(k)\right)u(k-i+1).
\end{equation}

In \eqref{eqn14}, $\mathbf{Y}= \begin{bmatrix} y(1) &y(2) &\cdots &y(N)\end{bmatrix}^{T}$, $\boldsymbol{\lambda}$  is the parameters set of the HMM, $\mathbf{P}= \begin{bmatrix} p(1) &p(2) &\cdots &p(N)\end{bmatrix}^{T}$, $\pi_{p(1)}$ is the initial state probability of scheduling variable at $k=1$, and $a_{p_{k-1}^{k}}$ is the transition probability from ${p(k-1)}$ to ${p(k)}$. The coefficients $g_i$s need to be predicted in \eqref{eqn14}. Hence, here we use polynomial functions to estimate $g_i$s, i.e. considering a $dth$-order polynomial function
\begin{equation}
\label{eqn16}
g_i(p(k))=h_{1i}{p(k)}^{d}+h_{2i}{p(k)}^{d-1}+\cdots+h_{(d+1)i},
\end{equation}
where $h_{li}$s are the coefficients of $g_i(p(k))$ in \eqref{eqn16}. For the sake of brevity, some new notations are defined as follows
\setlength{\arraycolsep}{0.0em}
\begin{eqnarray}
\label{eqn17}
&\phi_i{(k)}\overset{\Delta}{=} u(k-i+1),\nonumber\\
&\boldsymbol{\varphi}(k)\overset{\Delta}{=}\begin{bmatrix}\phi_{1}(k) &\cdots &\phi_{n_g}(k) \end{bmatrix}^{T},\nonumber\\
&\mathbf{P}_{k}^{d}\overset{\Delta}{=}\begin{bmatrix}{p(k)}^{d} &{p(k)}^{d-1} &\cdots &p(k) \end{bmatrix}^{T},\nonumber\\
&\mathbf{H}\overset{\Delta}{=}\begin{bmatrix}
h_{11} &h_{12} &\cdots &h_{1n_g}\\
h_{21} &h_{22} &\cdots &h_{2n_g}\\
\vdots &\vdots &\ddots &\vdots\\
h_{(d+1)1}&h_{(d+1)2} &\cdots &h_{(d+1)n_g}
\end{bmatrix},&
\end{eqnarray}
\setlength{\arraycolsep}{5pt}
thus, we can rewrite \eqref{eqn15} as
\begin{equation}
\label{eqn18}
\mu(k)=\mathbf{P}_{k}^{d}\mathbf{H}\boldsymbol{\varphi}(k).
\end{equation}
In \eqref{eqn18}, the estimation problem of the $g_i$s is converted to estimating the entries of matrix $\mathbf{H}$. Applying \eqref{eqn18}, the rewritten form of \eqref{eqn14} is
\begin{align}
    \label{eqn19}
Pr(\mathbf{Y},\mathbf{P}|\boldsymbol{\lambda})= \frac{\pi_{p(1)}}{{(2\pi {\sigma^2})}^{N/2}}\prod_{k=1}^{N-1}a_{p_{k-1}^{k}}\times\prod_{k=1}^{N}exp\left(\frac{-\left({y(k)-\mathbf{P}_{k}^{d}\mathbf{H}\boldsymbol{\varphi}(k)}\right)^2}{2\sigma^{2}}\right).
\end{align}
Since the entries of matrix $\mathbf{H}$ do not have impact on $Pr(\mathbf{P}|\boldsymbol{\lambda})$ , we can estimate them by maximizing \eqref{eqn19} with respect to matrix $\mathbf{H}$ as follows
\begin{equation}
\label{eqn20}
\mathbf{H}^{*}=arg\max_{\mathbf{H}}\left[exp(\frac{-\mathbf{J}}{2\sigma^{2}})\right],
\end{equation}
where
\begin{equation}
\mathbf{J}\overset{\Delta}{=}\sum_{k=1}^{N}\left({y(k)-\mathbf{P}_{k}^{d}\mathbf{H}\boldsymbol{\varphi}(k)}\right)^2.\nonumber
\end{equation}
In \ref{eqn20}, the optimum $\mathbf{H}$ is the matrix minimizing the amplitude of the $exp(.)$ function input argument. By differentiating of \eqref{eqn20} with respect to the entries of $\mathbf{H}$, the updating equation for each entry is calculated as follows
\begin{equation}
\label{eqn21}
h_{vw}^{*}=\frac{\sum_{k=1}^{N}\left[ \eta_{w}^{k}\left(p,\phi\right)\left(y(k)-\mathbf{P}_{k}^{d}\mathbf{H}^{-(vw)}\boldsymbol{\varphi}(k)\right)\right]}{\sum_{k=1}^{N}\left(\eta_{w}^{k}\left(p,\phi\right)\right)^{2}},
\end{equation}
where
\setlength{\arraycolsep}{0.0em}
\begin{eqnarray}
&\eta_{w}^{k}\left(p,\phi\right)=\left[p(k)\right]^{(d+1)-v}\phi_w(k),&\nonumber\\
&\left[\mathbf{H}^{-(vw)}\right]_{rs}=\left\{
\begin{array}{ll}
h_{rs};& r\neq v, s\neq w\\
0;& \mbox{elsewhere}\nonumber
\end{array}.
\right.&
\end{eqnarray}
\setlength{\arraycolsep}{5pt}
The main remaining problem is the estimation of the scheduling variable path. Here, we use the Viterbi algorithm to solve it. Using \eqref{eqn15}, the  variable $\mu_j(k)$ is defined as follows
\begin{equation}
\label{eqn22}
\mu_j(k)=\sum_{i=1}^{n_{g}}g_{i}\left(p(k)=p_j\right)u(k-i+1).
\end{equation}
Taking into account \eqref{eqn22}, we define the variable $\delta_j(k)$ in the Viterbi algorithm as follows
\begin{equation}
\label{eqn23}
\delta_j(k)=\frac{exp\left(\frac{-\left({y(1)-\mu_j(1)}\right)^2}{2\sigma^{2}}\right)}{\sigma \sqrt{2\pi}}\max_{1\leq i\leq M}{\delta_i(k-1)a_{ij}},
\end{equation}
where $\delta_j(1)$ in \eqref{eqn23} is defined as
\begin{equation}
\delta_j(1)=\frac{\pi_j }{\sigma \sqrt{2\pi}}exp\left(\frac{-\left({y(1)-\mu_j(1)}\right)^2}{2\sigma^{2}}\right).\nonumber
\end{equation}
In \eqref{eqn23}, the indices $1\leq i,j\leq M$ choose the corresponding state from $\mathbf{\mathcal{P}}= \begin{Bmatrix}p_1&p_2&\cdots& p_M\end{Bmatrix}$. Other parameters such as $\pi_j$s and $a_{ij}$s are the entries of the initial state vector $\boldsymbol{\pi}$ and the state transition matrix $\mathbf{A}$, respectively. Using the above definition in the Viterbi algorithm, the scheduling variable path is estimated by saving the vector of arguments that maximize \eqref{eqn23}. Other parameters such as $\pi_j$s and $a_{ij}$s are estimated by the Baum-Welch algorithm.

In the next section, the quasi--static Viterbi and Baum--Welch (QSVBW) algorithm are presented as an algorithm that can deal with the simultaneous estimation problem of the HMM parameters and the scheduling variable path in LPV context.

\section{The Proposed QSVBW Algorithm}\label{secIV}

In the previous section, the two main estimation problems of HMM are solved separately and it has been assumed that the required data for each estimation problem are available. Here, we need a static algorithm in order to estimate the scheduling variable, the entries of $\mathbf{H}$, and the HMM parameters. In this work, the QSVBW algorithm is devised to solve the aforementioned simultaneous estimation problem. The proposed QSVBW algorithm is as follows

\subsection{Use cross-validation to determine the order of polynomial models that approximate the coefficients of gis in an LPV static model.}\label{secIV.A}

\subsection{Determine the initial conditions:}\label{secIV.B}
\begin{itemize}
\item The transition matrix $\mathbf{A}$ and the initial state vector $\boldsymbol{\pi}$ using \eqref{eqn4} and \eqref{eqn6}. 
\item The entries of M by using an intelligent search algorithm such as particle swarm optimization (PSO), and genetic algorithm (GA).
\end{itemize}

\subsection{Estimate the best sequence of the scheduling variable using the Viterbi algorithm \eqref{eqn23}.}\label{secIV.C}

\subsection{Update the parameters of the model:}\label{secIV.D}
\begin{itemize}
\item The transition matrix and initial vector state using the Baum-Welch algorithm.
\item The entries of $\mathbf{H}$ by employing \eqref{eqn21}.
\end{itemize}

\subsection{Stop condition}\label{secIV.E}
\begin{itemize}
\item Increasing likelihood or the maximum iteration
\item If the stop condition is met, terminate the algorithm, otherwise go back to \ref{secIV.C}.
\end{itemize}

\begin{figure}[thpb]
      \centering
      \includegraphics[trim=0cm 1cm 010cm 0cm,scale=0.36]{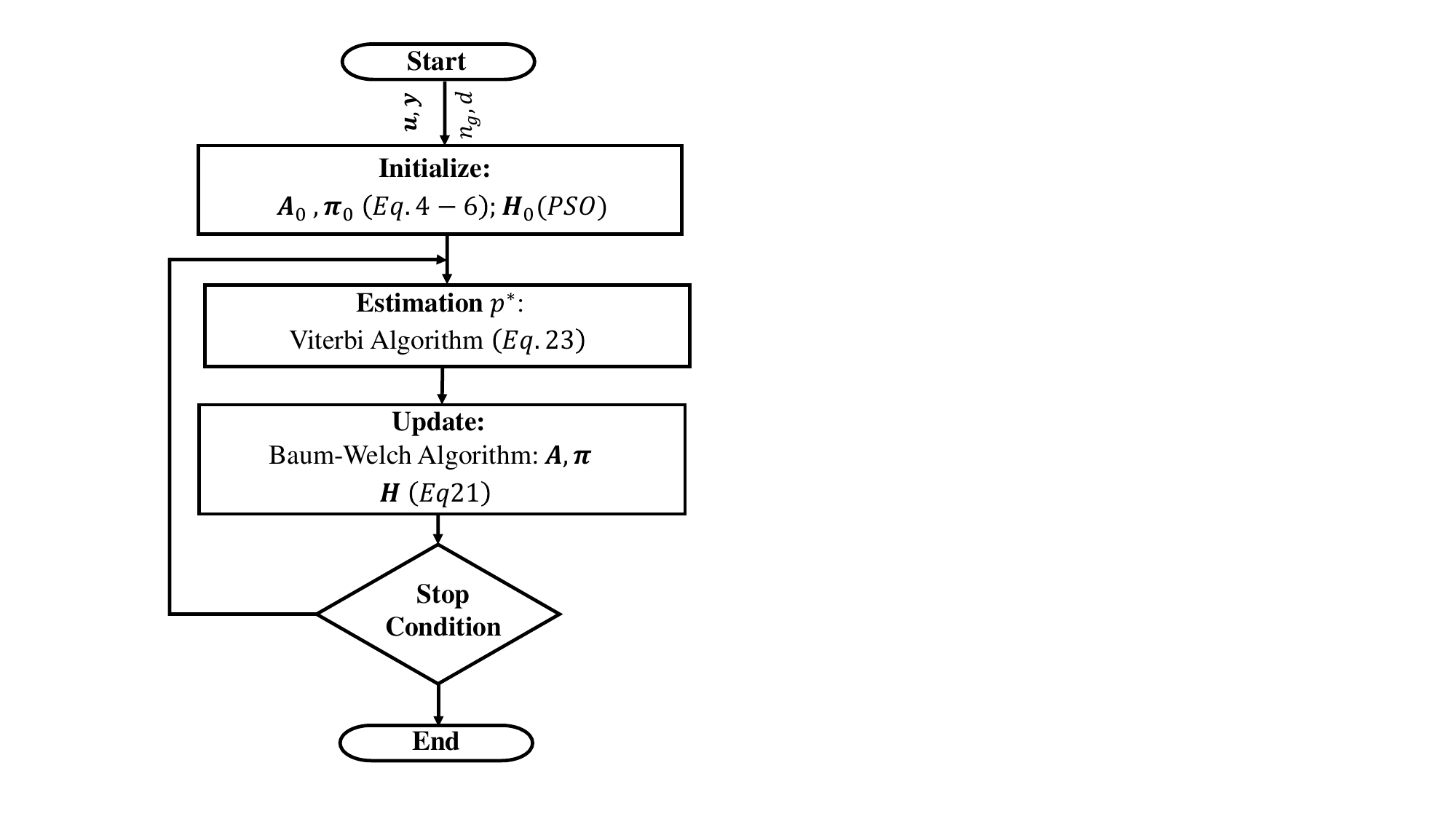}
      \caption{Flowchart of the QSVBW algorithm}
      \label{figlabel2}
   \end{figure}

In the QSVBW algorithm, the initial conditions for $\mathbf{A}$ and $\boldsymbol{\pi}$ are generated by considering \eqref{eqn4} and \eqref{eqn6}. The initial condition of $\mathbf{H}$ is generated by the PSO algorithm. The flowchart of the algorithm is shown in Fig. \ref{figlabel2}.

\section{Numerical Study}\label{secV}

To illustrate the performance of the QSVBW algorithm, an LPV-IO model with five nonlinear coefficients \cite{c18_golabi2014bayesian} is considered. Each of these coefficients has a different type of dependency with respect to the scheduling variable. The quality of the estimated signal will be assessed with the best-fit rate (BFR) criteria as follows
\begin{equation}
\label{eqn24}
BFR=100\%\times \max \left(0,1-\frac{{\left\lVert \mathbf{x}-\mathbf{\hat{x}} \right\rVert}_{2}}{{\left\lVert \mathbf{x}-\mathbf{\bar{x}} \right\rVert}_{2}}\right),
\end{equation}
where $\mathbf{x}$ is the vector of the true signal, $\hat{\mathbf{x}}$ is the vector of the estimated signal, $\bar{\mathbf{x}}$ is the mean value of the $\mathbf{x}$, $N$ is the size of the data, and $k$ denotes the associated sample.
The data generating system is defined as follows \cite{c18_golabi2014bayesian}
\begin{equation}
\label{eqn25}
y(k)=\sum_{i=1}^{5}g_i\left(p(k)\right)u(k-i+1)+e(k),
\end{equation}
where $\mathbf{\mathcal{P}}= \begin{Bmatrix}0.1&0.2&\cdots& 1\end{Bmatrix}$ and
\setlength{\arraycolsep}{0.0em}
\begin{eqnarray}
&g_1\left(p(k)\right)&=-exp\left(p(k)\right), g_2\left(p(k)\right)=1+p(k),\nonumber\\
&g_3\left(p(k)\right)&=\tan^{-1}\left(p(k)\right), g_4\left(p(k)\right)=-p(k),\nonumber\\
&g_5\left(p(k)\right)&=-\sin\left(p(k)\right).\nonumber
\end{eqnarray}
\setlength{\arraycolsep}{5pt}
A data set with 500 samples is generated by \eqref{eqn25}, and $u(k)$ is a periodic signal defined in \eqref{eqn26}.
\begin{equation}
\label{eqn26}
u(k)=\sin\left({\frac{2\pi k}{9}}\right).
\end{equation} 

The noise is assumed to be {\it{iid}} and has a Gaussian distribution. The signal-to-noise ratio (SNR) is set as $SNR=10 \log\left(\frac{P_y}{P_e}\right) =21.57$, where $P_y$ is the average power of the output signal, which is the deterministic component of $y$ in \eqref{eqn25}.
The data set is divided into train and test sets that have equal samples, each of them has 250 samples.  The train is used to extract the nonlinear mappings from inputs to output and the parameter of the equivalent HMM network of the corresponding LPV model and the test set is used to examine and verify the applicability and performance of the QSVBW algorithm.

\begin{figure}[thpb]
      \centering
      \includegraphics[trim=5cm 7.5cm 5cm 14.5cm, scale=0.68]{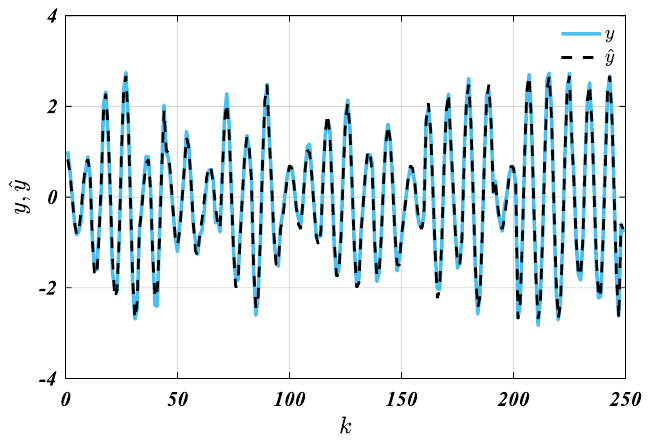}
           \caption{True output signal: $y$ (solid-blue), estimated output signal: $\hat{y}$ (dashed-black)}
      \label{figlabel3}
   \end{figure}

\begin{figure}[thpb]
      \centering
	 \includegraphics[trim=5.5cm 6.25cm 4.5cm 15.75cm, scale=0.68]{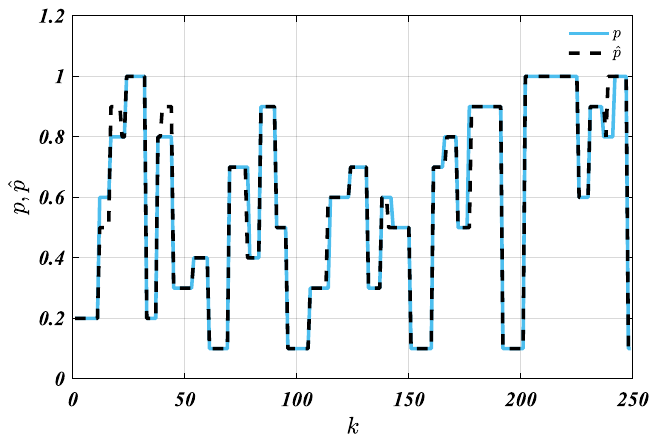}
      \caption{True scheduling variable: $p$ (solid-blue), estimated scheduling variable: $\hat{p}$ (dashed-black)}
      \label{figlabel4}
   \end{figure}

The true output signal and its estimated signal are shown in Fig. \ref{figlabel3}. Regarding \eqref{eqn24}, the fitness score is 95.67, and it is shown that the proposed QSVBW algorithm approximately extracts the nonlinear mappings of the system. This score belongs to four $4^{th}$-order polynomials, $\mathbf{H}$ is a 5 by 4 matrix, as the best approximator of the system, described in \eqref{eqn25}. The estimated scheduling variable and the true scheduling variable signal are indicated in Fig. \ref{figlabel4}, and the fitness score is 90.26. Fig. \ref{figlabel4} shows that the proposed algorithm can perfectly estimate the scheduling variable, as the unknown input in the BLPVI problem. 

\section{Discussion}\label{secVI}
The proposed QSVBW algorithm estimates scheduling variables in LPV models with static dependency. Based on the dynamic dependency on scheduling variables, the first-order Markov chain is not recommended and high-order Markov chain could be implemented in a hidden state layer. Since the Viterbi algorithm estimates the scheduling variable based on the last state at each iteration, the proposed QSVBW algorithm needs some modification in the estimation step to deal with dynamic dependency in the coefficients. Therefore, here we acclaim that the proposed QSVBW algorithm is just applicable when the coefficients have a static dependency on the scheduling variable, and it is not capable of handling the problem when nonlinear mappings have a dynamic dependency on the scheduling variable.

The system is assumed to be SISO while this is not necessary and only used to simplify the mathematical equations. In the case of multi-input multi-output (MIMO) systems, a D-dimensional Gaussian pdf is allocated to the observations of the system. The system is then stated in a HMM framework and the best sequence of scheduling variables is estimated by the proposed QSVBW algorithm with no loss of generality.

The proposed QSVBW algorithm uses polynomial functions to approximate the nonlinear maps in \eqref{eqn16}. Such a choice enables us to reach a closed-form mathematical solution with respect to the problem of approximating nonlinear coefficients. Spline functions are piecewise polynomial curves, and they are another choice to approximate the nonlinear maps in a closed--form solution. In approximation of the nonlinear maps, other methods like neural networks and fuzzy models do not reach a closed-form solution; hence, the problem is more complex in this sense. In the QSVBW algorithm, a considerable issue is finding a proper initial condition, especially for the coefficient matrix, $\mathbf{H}$. The size of $\mathbf{H}$ obtained in numerical studies is 5 by 4, four $4^{th}$-order polynomials; thus, finding a suitable initial condition is a severe challenge in this very large search--space. To solve this problem, the PSO algorithm is used. The PSO algorithm is an intelligent search method that looks for the optimum solution locally and globally between its particles. Although it is a time-consuming procedure, it is faster than other search methods GA in large search spaces.

\section{Conclusion}\label{secVII}
In this paper, a new blind system identification problem called BLPVI in LPV-IO structure is introduced. The unknown input signal in BLPVI is the scheduling variable signal. The scheduling variable and nonlinear functions are estimated by the proposed QSVBW algorithm. The QSVBW algorithm is a novel procedure that empowers us to express a static LPV model like a discrete-time HMM, and we then can use the characteristics of an HMM network to predict the scheduling variable signal. The numerical study showed that the proposed algorithm is capable of estimating scheduling variables and nonlinear coefficients perfectly, and it is not sensitive to the type of nonlinearity in coefficients. Our feature work deals with solving blind system identification problems where the coefficients have a dynamic dependency on the scheduling variable.

\section*{APPENDIX} \label{appendix}
In this section, we present the proof of the updating criteria obtained in \eqref{eqn21}. Considering \eqref{eqn20}, the minimizer of $\mathbf{J}$ is the optimal $\mathbf{H}^{*}$. One can rewrite $\mathbf{J}$ as follows
\begin{align*}
  \mathbf{J}=&\sum_{k=1}^{N}\bigg[y(k)-\left(h_{11}[p(k)]^{d}+h_{2}[p(k)]^{d-1}+\cdots + h_{n_{g}1}\right)\phi_1{(k)} +\nonumber\\
  &\vdots\\
&+\left(h_{1n_{g}}[p(k)]^{d}+h_{2n_{g}}[p(k)]^{d}+\cdots +h_{(d+1)n_{g}}\right)\phi_{n_g}{(k)}\bigg]^2,\nonumber  
\end{align*}
therefore, by differentiating of $\mathbf{J}$ with respect to the each entry of $\mathbf{H}$, one will obtain 
\begin{align*}
&\frac{\partial\mathbf{J}}{\partial h_{vw}}=2\sum_{k=1}^{N} [p(k)]^{d+1-v}\phi_w(k)\left({y(k)-\mathbf{P}_{k}^{d}\mathbf{H}\boldsymbol{\varphi}(k)}\right)\nonumber\\
&\sum_{k=1}^{N} [p(k)]^{d+1-v}\phi_w(k)y(k)=\sum_{k=1}^{N} [p(k)]^{d+1-v}\phi_w(k)\mathbf{P}_{k}^{d}\mathbf{H}\boldsymbol{\varphi}(k).\nonumber
\end{align*}
Considering the definition of $\eta_{w}^{k}(p,\phi)$ and $\mathbf{H}^{-(vw)}$ in \eqref{eqn21}, then
\begin{equation}
\label{eqn_appendix}
\sum_{k=1}^{N}\eta_{w}^{k}(p,\phi)\left(y(k)-\mathbf{P}_{k}^{d}\mathbf{H}^{-(vw)}\boldsymbol{\varphi}(k)\right)=h_{vw}^{*}\sum_{k=1}^{N}\left[\eta_{w}^{k}(p,\phi)\right]^2 \nonumber
\end{equation}
Finally, $h_{vw}^{*}$ will be calculated by \eqref{eqn21}.


\bibliographystyle{IEEEtran}
\balance
\bibliography{Ref.bib}
\end{document}